\documentclass[runningheads]{llncs}
\usepackage{graphicx}
\usepackage{amsmath}
\usepackage{amsfonts}
\usepackage{xcolor}
\usepackage{color}
\usepackage{makecell}

\begin{document}
\title{Migration-Related Semantic Concepts for the Retrieval of Relevant Video Content}
%
%
 \author{Erick Elejalde\inst{1}  \and
Damianos Galanopoulos \inst{2} \and Claudia Nieder\'ee\inst{1}
 \and Vasileios Mezaris \inst{2}}

\institute{L3S Research Center,
Leibniz-University, Hannover,
Germany \\ \email{\{elejalde,niederee\}@l3s.de} \and
CERTH-ITI, Thermi-Thessaloniki, Greece\\ \email{\{dgalanop,bmezaris\}@iti.gr} }
\authorrunning{E. Elejalde et al.}
\titlerunning{MRSCs for the retrieval of relevant video content} 

%
\maketitle              
\begin{abstract}
Migration, and especially irregular migration, is a critical issue for border agencies and society in general. Migration-related situations and decisions are influenced by various factors, including the perceptions about migration routes and target countries. An improved understanding of such factors can be achieved by systematic automated analyses of media and social media channels, and the videos and images published in them. However, the multifaceted nature of migration and the variety of ways migration-related aspects are expressed in images and videos make the finding and automated analysis of migration-related multimedia content a challenging task. We propose a novel approach that effectively bridges the gap between a substantiated domain understanding - encapsulated into a set of Migration-related semantic concepts - and the expression of such concepts in a video, by introducing an advanced video analysis and retrieval method for this purpose.

\end{abstract}
\section{Introduction}
Migration is a complex process, where decisions are driven by a multitude of factors, including the perceptions about migration routes and target countries~\cite{Timmermann2014}.  Media, and especially social media, with their powerful use of images and videos from multiple sources, plays an essential role in forming and manipulating such perceptions and misperceptions \cite{Bakewell2016}, e.g., via misinformation campaigns.  A better understanding of the media and its impact, thus, can help in anticipating possible migration-related risks at the border and in transit countries. 

With the vast amounts of media and media channels from a wide variety of sources, automated content analysis is necessary. However, the multifaceted nature of migration and the range of ways related aspects are expressed in images and videos make the finding and automated analysis of migration-related content very challenging.

To address some of these challenges, we propose a novel method, which effectively combines a top-down with a bottom-up approach. We leverage the substantial theoretical understanding that has been achieved on migration factors and migration decisions~\cite{doi:10.1177/030913259501900404,10.2307/2546431,Sassen1988}. For this, we define a domain conceptualization in close collaboration with experts, resulting in a set of Migration-Related Semantic Concepts (top-down). This is combined with an advanced video analysis method that captures and retrieves the different ways these semantic concepts are expressed in videos and images (bottom-up). Since the Migration-Related Semantic Concepts (MRSCs) are often abstract definitions (e.g. `\textit{ethnic identity}', `\textit{law enforcement}', etc.), it is very challenging to find and annotate video exemplars with these concepts.

Typical concept annotation and retrieval methods use image/video exemplars as training materials to develop pre-defined concept detectors \cite{certh2015,MarkatopoulouMM6,certh2013}. However, these methods suffer from scalability limitations because it is difficult to collect and annotate large enough datasets. Moreover, it is very time and effort consuming to integrate new concepts due to the manual annotation and training phases. To overcome these limitations, we aim at associating MRSCs with visual content without using any training visual exemplars. We adopt a state-of-the-art approach for Ad-hoc Video Search (AVS) that directly transforms visual and textual content into a common feature space, in which a straightforward comparison is feasible. AVS is a type of cross-modal retrieval problem, in which video shots are recovered when the query is a complex textual sentence. Similarly, our MSRCs retrieval problem is to identify and annotate images or video shots with MRSCs, starting from the textual definition of the MRSCs. 

Early attempts on the AVS problem have relied on large sets of pre-trained visual concept detectors and NLP techniques for query analysis to find relevant visual concepts in the query. In  \cite{markatopoulou2017query}, the association between visual concepts and the textual queries was reached by using complex NLP rules and a vast set of pre-trained deep neural networks for video annotation. More recently, the problem has been addressed using deep neural networks to transform both the textual queries and the visual content in a new shared space \cite{Video2vec}. The dual encoding network proposed in \cite{dong2019dual} uses multiple levels of encoding to transform videos and queries into a common dense representation using an improved loss function \cite{faghri2018vse++}. An extension of the above was presented in \cite{galanopoulos2020}, where state-of-the-art results were achieved using rich representations and attention-based layers of encoding for both the text and the visual modalities. We build on this method to address the identification of visual content that could be associated with MRSCs.

\section{Migration-Related Semantic Concepts (MRSCs)}
A semantic concept is understood here as a meaningful entity or a comprehensive idea formed in the person's mind from the information perceived and the person's background. Moreover, a semantic concept is intrinsically linked to a context~\cite{eriksson2010concept}. Based on an in-depth study of migration theories and discussions with domain experts, we have collected semantic concepts that will help specialists to express the migration aspects they are interested in and to identify relevant visual information to be later interpreted and analyzed.

The collection of concepts has resulted in a set of 106 Migration-Related Semantic Concepts (MRSCs), i.e., concepts relevant in the context of migration. These are organized in five general categories (see Section \ref{sec:factors_classification}).

The advantages of basing our video retrieval process in MRSCs are many-fold. First, it provides a substantiated and diverse common language (with ground definitions of concepts), which can be used to express information needs; and be used in later expert analysis of the retrieved information. Also, by critically relying on theoretically founded categories and discourses of migration, we minimize distorted or biased views of the subject under study and foster the consideration of a wide variety of aspects.

\subsection{Migration Theories}
We base our analysis mostly on three popular theoretical approaches used to study migration-related issues: the Neo‐classical economic equilibrium perspective, Historical‐structural approach, and Migration systems theory~\cite{de2019age,doi:10.1080/1369183X.2017.1384135}. We took these models into account to create a hierarchical definition of the concepts. However, as these models (for most purposes) are orthogonal to each other, our concepts and their classification are not tied to one particular theory. Instead, in combination, they help us to better identify concepts in the context of migration.

In essence, the \textit{Neo‐classical theory} centers on the imbalanced conditions between the country/region of origin and that of destination. It assumes that individuals in the process of migration will try to maximize their benefit in composition with other external constraints. Alternatives are compared, and from the gathered information, the potential migrant will decide if it is more `profitable' to stay or leave~\cite{10.2307/2546424}.
Related to the Neo-classical theory is the \textit{Push‐Pull framework}~\cite{10.2307/2060063}, which continues to emphasize the drivers of the flow of people. The underlying assumption of this theory is that negative factors at the origin push people away, while positive factors at the destination pull people toward them. The Push‐Pull model has been adapted and extended in various ways~\cite{doi:10.1080/1369183X.2017.1384135}. For example, in~\cite{doi:10.1177/030913259501900404}, the author suggests to include the mooring dimension (referred to as the Push‐Pull‐Mooring (PPM) theory).  The mooring factors are equivalent to moderating variables in that they can either potentiate migration or dissuade the migrants from leaving their current country.

The \textit{historical‐structural approach} provides an alternative to explain the migration processes. It also stresses the unequal distribution of economic and political influence in the world's economy but is mostly based on the Marxist view of the political economy \cite{robincohen1987,Sassen1988}. 
The historical-structuralist accounts shift the focus away from voluntary migration (as suggested by push-pull models) to a global scale recruitment of labor by capital~\cite{de2019age}. Here, migration is presented as a means of mobilizing cheap labor for capital, which, in turn, preserves the existing uneven development. 

Both the neo-classical perspective and the historical-structuralist approach have faced criticism~\cite{10.1525/j.ctt7zw0nw,hein2011a,10.2307/2546431}. The former is accused of overlooking historical antecedents of movements, and underestimating the role of the state, while the latter attributes to the interests of capital most of the weight in the migration process, and pays almost no attention to the personal motivations~\cite{de2019age}. As a response to these criticisms, comes up the third model considered in our study, i.e., \textit{Migration systems}~\cite{10.2307/2546431,castelli2018drivers}. The migration systems approach proposes a more holistic analysis that examines the origin and the destination by considering all the linkages between the two places. 
This approach suggests that the migratory process can be represented as the result of interacting macro‐, meso- and microstructures~\cite{castelli2018drivers}. Macro‐structures are represented by large institutional factors (mostly out of the control and independent from the migrant - e.g., the political economy of the world market). Micro-structures are understood as the pattern of relations between essential elements of the social life that cannot be further divided and have no social structure of their own (e.g., cultural capital).
Finally, meso-structures are located in the space between micro and macro-structures. These act as obstacles or facilitators in the migratory process. Even when they are related to the individual, they are not entirely under her/his control (e.g., technology, migration industry, etc.). All these structures are interconnected and help to describe the entire process from the migration decision to settlement and community formation~\cite{de2019age}.

\subsection{Factors Classification}
\label{sec:factors_classification}
Semantic concepts can be combined to form meaningful templates, containing several aspects and, in turn, specifying further semantics. For example, the aspects `family' and `war' can be combined in a template as `Families in war'. These templates can then also be used as a semantic concept. Based on these patterns, and similar to comparable works such as the World bank theme taxonomy\footnote{\url{http://pubdocs.worldbank.org/en/275841490966525495/Theme-Taxonomy-and-definitions.pdf}}, we have defined the MRSCs as a hierarchical structure. This tree-style classification further allows annotating information at different aggregation levels and contributes to the definition of each concept's context.

\begin{table}
\centering
\caption{Dataset of MRSCs organized in five general categories and in two hierarchical levels. In the second level column we are including only a subset of the complete list.}
\label{tab:mrsc}
\resizebox{0.99\textwidth}{!}{
\begin{tabular}{l|l|l}
\hline
\textbf{Category}    &   $\mathbf{1^{st}\ Level}$  &   $\mathbf{2^{nd}\ Level\ (examples)}$ \\
\hline
Economic            & Labour market           & Working conditions, Labor movements, Job segmentation \\
                    & Migrant groups infrastructure &  Migration industries, Family labor, Socio-spatial texture \\
                    & Capital flows           & Investment, Informal economic activities, International trade \\
\hline
Social              & Ethnic minority formation & Others-definition, Self-definition, Ethnic identity \\
                    & Ethnic community formation & Cohesion, Access to community, Access to information and services \\
                    & Ethnicity                 & Xenophobia, Racism, Language \\
                    & Interpersonal relationships & Radicalization, Rumors, Cultural interaction \\
                    & Cultural capital           & Adaptability, \textbf{\textit{Education}}, Knowledge of other country \\
                    & Social capital             & Informal social activities, Opinion formers \\
\hline
Demographic         & Target-earners             & Remittances, Relative success/failure in target country \\
                    & Gender                     & Marriage, Domestic service, Caretaking \\
	            & Skilled professionals      & Brain drain \\
                    & Refugees                   & \textbf{\textit{War}}, \textbf{\textit{Political instability}}, Persecution \\
\hline
Environmental       & \textbf{\textit{Urbanization}}     & Global cities, Stopgaps, Ethnic footholds, \textbf{\textit{Access to medical care}} \\
                    & Ecology                   & Climate change, Pollution, Natural disasters \\
\hline
Political           & Settlement                & Citizenship, Laws, Nation \\
                    & Immigration policies      & Representation of immigrants in policies, Change of policies over time \\
                    & Crime                     & Acculturation problems, Ethnic tensions, Cultural predispositions \\
                    & Organized crime           & Human trafficking, Document fraud, Money laundering \\
                    & Politics                  & Corruption, Policymakers, Regulatory hoops/ their avoidance \\
\hline
\end{tabular}}
\end{table}

The identified concepts are grouped into five categories: economic, social, demographic, environmental, and political. These classes are consistent across the migration literature~\cite{castelli2018drivers,de2019age,10.2307/2546431}. We identified 106 MRSCs that we organized on two levels: 20 on the first level and the other 86 under them (see Table \ref{tab:mrsc}).

As expected, many social factors are related to migration. 
One crucial definition in this category is that of `Ethnicity'. The concept of ethnicity is relevant in more than one of the sub-classes. However, we have specifically identified a class called `Ethnicity', which refers to a real process of historical individuation by linguistic and cultural practices that give a sense of collective identity~\cite{cohen1988multi}.

Another important category identified concerning MRSCs is `Demographic'. This category comprises concepts such as `Target‐earners', `Gender', and `Refugees'. Identifying gender‐specific perceptions and expectations during the migration process is critical in the analysis of media content. Many gender‐related misperceptions lead to conflicts and security issues\footnote{\url{https://migrationdataportal.org/themes/gender-and-migration}}. This is why one essential class of issues to be analyzed relates explicitly to `Gender'. This group of concepts will deal with terms such as `Marriage', `Care-taking', and `Domestic service' (all of which are disproportionately associated with women). Another relevant concept in this category is `Refugees'. This, in turn, can be contextualized in terms of more specific concepts such as `War' or `Political instability', which are identified as factors influencing forced migration. 

Environmental factors are becoming more and more relevant when it comes to migration. Environmental migrants or climate refugees are forced to leave their home region due to sudden or long‐term changes to their local environment. These changes compromise their well‐being or their secure livelihood. Such changes may include increased droughts, desertification, sea‐level rise, and disruption of seasonal weather patterns. Other environmental factors that bring people to migrate are those related to an increasing tendency toward urbanization. This includes topics such as `Global cities', `Stopgaps', `Ethnic footholds', and other factors like better `access to medical care' that drive migrants looking to improve their quality of life.

It is crucial to notice that our list of MRSCs is defined based on the specific interests of our current research, and they do not constitute a comprehensive list of all the concepts related to migration. Also, despite the tree-style hierarchical classification of the concepts, these are not necessarily mutually exclusive. For example, `remittances' is a concept that can be associated with economic factors such as `Capital flow', but also is strongly related to demographic factors as one of the main drivers for `Target-earners' (i.e., economically active people who want to save enough in a higher-wage economy to improve conditions at home).

\section{MRSC-Based Video Retrieval}
\label{sec:MRSC-based_video_retrieval}
 
\begin{figure}
  \centering
  \includegraphics[width=\linewidth]{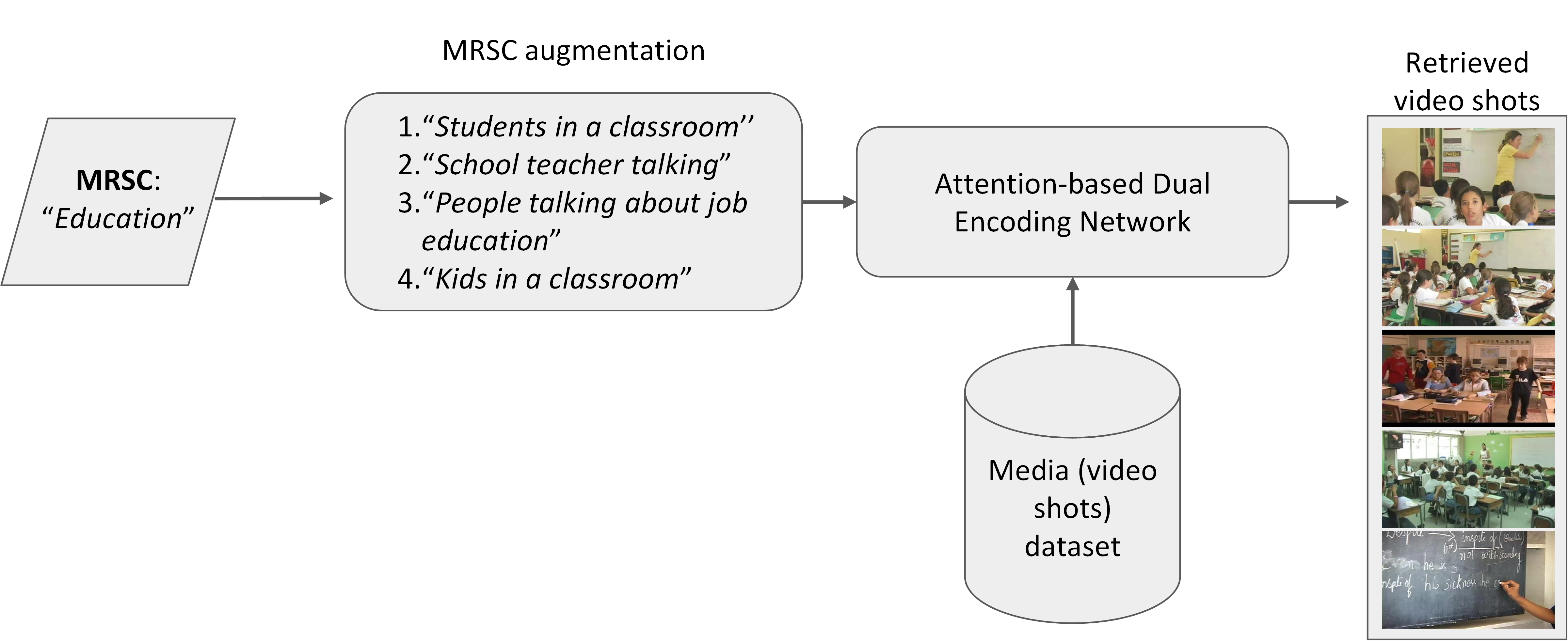}
  \caption{An overview of the MRSC retrieval method}
 \label{mrsc_fig:1}
\end{figure}

We address the MRSC-based video retrieval problem by adapting a previously developed state-of-the-art method for the AVS problem \cite{galanopoulos2020}. This approach's overall idea is to train a deep neural network (DNN) by using video-caption pairs. This DNN is then used as a video retrieval system by inputting MRSCs to recover the most related video shots. Since the MRSCs are typically high-level abstractions of concepts, we choose to enrich them so that we have an information-rich input to the video retrieval system. For this, each MRSC is manually complemented (i.e., augmented) with a small set of complex sentences that describe it. For instance, for the MRSC `\textit{Education}', sentences like ``\textit{students in a classroom attend a lecture}'' are added. This approach is closely related to the training procedure for our network. The MRSC (with its descriptions) and video shots from the target dataset are used as input to our system. They are encoded into the common feature space, and for every MRSC, a ranked list with the most related media items within the given image/video dataset is generated. An overview of the proposed method is illustrated in Figure \ref{mrsc_fig:1}.
 
For MRSCs-related video retrieval, we adjust the attention-based dual encoding network presented in \cite{galanopoulos2020}. This network utilizes two similar modules, each consisting of multiple encoding levels, for the visual and textual content respectively, along with a text-based attention component for more efficient textual representation. The network translates a media item (e.g. an entire video or a video shot) $\mathbf{V}$ and a textual item (e.g. a video shot caption or a text query) $\mathbf{S}$  into a new shared feature space $\phi(\cdot)$, resulting in two new representations $\phi(\mathbf{V})$ and $\phi(\mathbf{S})$ that are directly comparable. 

More specifically, each video shot is encoded into a three-level representation $ [\phi(V_1),\phi(V_2),\phi(V_3)]$. Firstly, the video shot is decoded into a fixed number of $n$ keyframes and fed into a pre-trained DCNN, from where a feature vector $v_n$ is produced for every keyframe. $\mathbf{V} = \{v_1,v_2,\dotsc,v_n\}$ is the collection of keyframe feature vectors for the shot and $\phi(V_1)=\frac{1}{n}\sum_{i=1}^{n}v_i$ is considered as the first-level video representation. The keyframes vectors are fed in a sequence of bi-directional Gated Recurrent Units (bi-GRUs) \cite{cho2014learning}, and their output $\mathbf{{H_v}} \in \mathbb{R}^{n\times2*h}$, where $h$ is the output size of a GRU cell, is forwarded into a self-attention mechanism \cite{galanopoulos2020} resulting in a weight matrix $\mathbf{A_v} \in \mathbb{R}^{n\times n}$. Then, the matrix $\mathbf{\overline{H}_v} = \mathbf{A_v}\mathbf{H_v} $ is calculated.
 $\phi(V_2)$ is calculated as $\phi(V_2)= \frac{1}{n}\sum_{t=1}^{n} \overline{h}_t$, where $\overline{h}_t$ is the $t^{th}$ row of $\mathbf{\overline{H}_v}$.
Finally, $\mathbf{\overline{H}_v}$ is forwarded into a  1-d convolutional layer resulting in the third-level representation $\phi(V_3)$ of the shot. The overall video shot representation is the concatenation of these three representations, which is forwarded into a trainable fully connected layer.
 
\begin{figure}
  \centering
  \includegraphics[width=\linewidth]{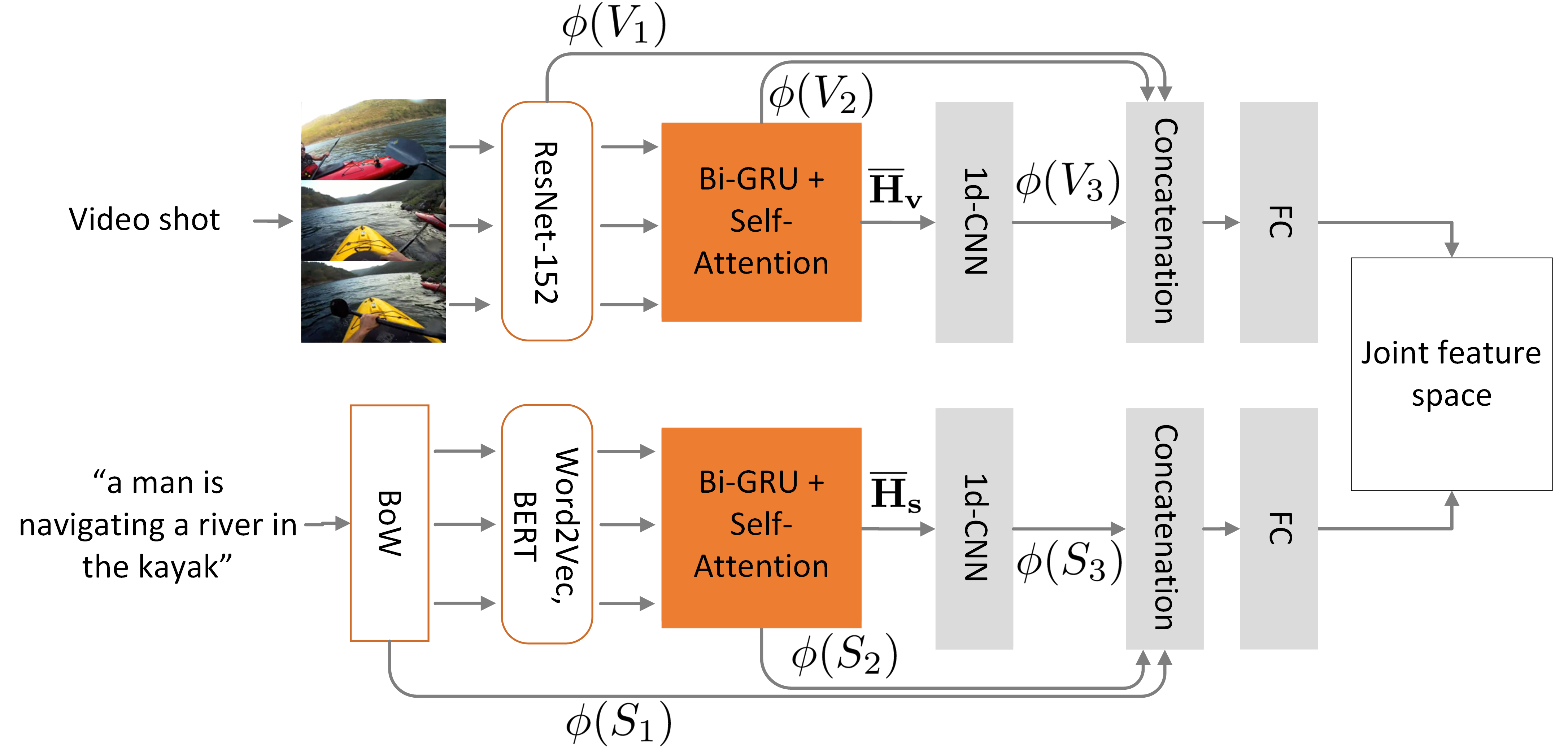}
  \caption{An overview of the attention-based dual encoding network}
 \label{mrsc_fig:2}
\end{figure}

Similar to the visual encoding module, a three-level representation  $[\phi(S_1)$, $\phi(S_2)$, $\phi(S_3)]$ is built for every textual item $S$. Considering a text sentence  $S$ as a set of $m$ words $\mathbf{S} = \{w_1,w_2,\dotsc,w_m\}$, the first-level $\phi(S_1)$ is created by averaging individual one-hot-vectors of these words. Then, for every word, a deep network-based word embedding vector $c_m$ is created, and is used as input for the bi-directional GRU module. The output of GRU $\mathbf{{H_s}} \in \mathbb{R}^{m\times2*h}$, is forwarded into the text-based attention mechanism resulting in a matrix $\mathbf{\overline{H}_s}= \mathbf{A_s}\mathbf{H_s}$, where $\mathbf{A_s} \in \mathbb{R}^{m\times m}$. Analogous to $\phi(V_2)$,  $\phi(S_2)= \frac{1}{m}\sum_{t=1}^{m} \overline{h}_t$, where $\overline{h}_t$ is the $t^{th}$ row of $\mathbf{\overline{H}_s}$. Finally, $\mathbf{\overline{H}_s}$ is fed to a 1-d convolutional layer resulting in the textual third-level representation  $\phi(S_3)$. Similar to the visual module, the overall textual representation is the concatenation of these three representations, which is forwarded into a trainable fully connected layer. Following the state of the art approach \cite{dong2019dual,faghri2018vse++,galanopoulos2020}, the improved marginal ranking loss is used to train the entire network.  The overview of the adapted attention-based dual encoding network is illustrated in Figure \ref{mrsc_fig:2}.
 
\section{Experiments and Results}
To train our network, we used the combination of two large-scale video datasets: MSR-VTT \cite{xu2016msr} and TGIF \cite{li2016tgif}. As initial keyframe representations, we use a ResNet-152 (trained on the ImageNet-11k dataset). Also, two different word embeddings are utilized: i) the Word2Vec model \cite{mikolov2013efficient} trained on the English tags of 30K Flickr images, provided by \cite{dong2018predicting}; and, ii) the pre-trained language representation BERT \cite{devlin2018bert}, trained on Wikipedia content. To evaluate the performance of our network for MRSCs retrieval, since there is no available dataset for this, we use the evaluation datasets of the TRECVID Semantic Indexing task (SIN) for the years of 2013 and 2015\footnote{\url{https://www-nlpir.nist.gov/projects/tv2015/index.html\#sin}}. The goal is to retrieve the most related video shots by inputting the names of a set of visual concepts. In our analysis, these concepts take over the position of the MRSCs for evaluation purposes. A good performance on these datasets will document the merit of the proposed approach for the needs of the MRSC-based retrieval problem. The mean extended inferred average precision (MXinfAP) is used as an evaluation measure, as is usually the case with these datasets \cite{over2013trecvid}.

\begin{table}
\centering
\caption{Results of video shot retrieval in the SIN'13 and SIN'15 datasets for 38 and 30 visual concepts, respectively, in terms of XinfAP. The ``Concept name'' column presents the results when we use as textual input only the concept label. In contrast, the ``Concept name + descriptions'' column stands for the setup in which the concept label is augmented with short sentences describing the concept further.}
\label{mrsc_tbl:1}
\resizebox{0.99\textwidth}{!}{
\begin{tabular}{l|cc||l|cc}
\hline \multicolumn{3}{c||}{SIN'13 dataset} & \multicolumn{3}{c}{SIN'15 dataset} \\ \hline
    & \parbox{1.3cm} {Concept name} & \parbox{1.8cm} {Concept name + descriptions} &    & \parbox{1.3cm} {Concept name} & \parbox{1.8cm} {Concept name + descriptions}\\ \hline
1003 Airplane & 0.1928 & 0.2789 & 1003 Airplane & 0.3254 & 0.5055 \\
1005 Anchorperson & 0.0128 & 0.0646 & 1005 Anchorperson & 0.0067 & 0.0145\\
1006 Animal & 0.0253 & 0.1748 & 1009 Basketball & 0.0134 & 0.1814\\
1010 Beach & 0.4648 & 0.515 & 1013 Bicycling & 0.0569 & 0.373\\
1015 Boat Ship & 0.3653 & 0.4443 & 1015 Boat Ship & 0.4804 & 0.5998 \\
1016 Boy & 0.0601 & 0.1279 & 1017 Bridges & 0.085 & 0.1615\\
1017 Bridges & 0.0268 & 0.0688 & 1019 Bus & 0.1215 & 0.1382\\
1019 Bus & 0.0657 & 0.112 & 1022 Car Racing & 0 & 0.0647\\
1025 Chair & 0.0309 & 0.1207 & 1027 Cheering & 0.0004 & 0.0687 \\
1031 Computers & 0.112 & 0.2982 & 1031 Computers & 0.148 & 0.362 \\ \hline
1038 Dancing & 0.0242 & 0.1503 & 1038 Dancing & 0.0002 & 0.1239\\
1049 Explosion Fire & 0.1884 & 0.2582 & 1041 Demonstration Or Protest & 0 & 0.2574\\
1052 Female Human Face Closeup & 0.1017 & 0.1459 & 1049 Explosion Fire & 0.104 & 0.1739\\
1053 Flowers & 0.1035 & 0.1661 & 1056 Government Leader & 0.0003 & 0.1677\\
1054 Girl & 0.0388 & 0.1271 & 1071 Instrumental Musician & 0.0002 & 0.3458\\
1056 Government Leader & 0 & 0.2767 & 1072 Kitchen & 0.0805 & 0.34\\
1059 Hand & 0.0904 & 0.1025 & 1080 Motorcycle & 0.1303 & 0.236\\
1071 Instrumental Musician & 0.0031 & 0.3305 & 1085 Office & 0.0546 & 0.2425 \\
1072 Kitchen & 0.0745 & 0.1537 & 1086 Old People & 0.0473 & 0.1993\\
1080 Motorcycle & 0.2042 & 0.2581 & 1095 Press Conference & 0.0001 & 0.0219\\ \hline
1083 News Studio & 0.0206 & 0.0609 & 1100 Running & 0.0008 & 0.0178\\
1086 Old People & 0.0854 & 0.2108 & 1117 Telephones & 0 & 0.3088 \\
1089 People Marching & 0 & 0.0626 & 1120 Throwing & 0.0001 & 0.0485\\
1100 Running & 0.0059 & 0.1494 & 1261 Flags & 0.0685 & 0.156\\
1105 Singing & 0.0008 & 0.1057 & 1297 Hill & 0.0319 & 0.0675\\
1107 Sitting Down & 0.0001 & 0.0084 & 1321 Lakes & 0.0577 & 0.2033 \\
1117 Telephones & 0 & 0.3151 & 1392 Quadruped & 0.0017 & 0.2311 \\
1120 Throwing & 0 & 0.125 & 1440 Soldiers & 0.2436 & 0.3709\\
1163 Baby & 0.2991 & 0.4707 & 1454 Studio With Anchorperson & 0.0021 & 0.0393\\
1227 Door Opening & 0.0177 & 0.0377 & 1478 Traffic & 0.1372 & 0.2046 \\ \hline
1254 Fields & 0.0192 & 0.1578 &  &  &  \\
1261 Flags & 0.1274 & 0.2687 &  &  &  \\
1267 Forest & 0.1026 & 0.1939 &  &  &  \\
1274 George Bush & 0 & 0.44 &  &  &  \\
1342 Military Airplane & 0.0001 & 0.1062&  &  &  \\
1392 Quadruped & 0.0214 & 0.2928 &  &  &  \\
1431 Skating & 0.2684 & 0.424 &  &  &  \\
1454 Studio With Anchorperson & 0.0047 & 0.0419 &  &  &  \\ \hline
Mean XinfAP & 0.0831 & \textbf{0.2012} &  & 0.0733 & \textbf{0.2075} \\ \hline
\end{tabular}}
\end{table}

\begin{table}
\centering
\caption{Comparison with SIN Task-specific methods in the SIN'13 and SIN'15 datasets, in terms of XinfAP.}
\label{mrsc_tbl:2}
\begin{tabular}{l|c|c}
     & \multicolumn{1}{l|}{SIN 2013} & \multicolumn{1}{l}{SIN 2015} \\ \hline
\begin{tabular}[l]{@{}l@{}}Proposed AVS approach\\ (concept name + descriptions; no training exemplars)\end{tabular}                  & 0.2012                    & 0.2075                     \\
\cite{MarkatopoulouMM6} (using annotated exemplars for training) & 0.2504                        & -                            \\
\cite{certh2013} (using annotated exemplars for training)& 0.1580                        & -                            \\
\cite{certh2015}  (using annotated exemplars for training)            & -                             & 0.263                       \\ \hline
\end{tabular}
\end{table}

We compare our proposed model with conventional concept retrieval methods that use predefined sets of visual concepts, positive exemplars for every concept, and are trained on these sets. The goal is to highlight the performance of our approach that does not require concept-annotated training videos. This gives our model a practical advantage over the supervised learning methods. 

Table \ref{mrsc_tbl:1} presents the results on the SIN'13 and SIN'15 datasets for the detection of 38 and 30 different concepts, respectively. The results of Table \ref{mrsc_tbl:1} show that the use of additional information (i.e., ``augmentations'') for every concept (see Section \ref{sec:MRSC-based_video_retrieval}) leads to a significantly improved performance. Examples of substantial improvements are the concepts \textit{``Telephones''} in both datasets, and {\textit{``Bicycling''}} and \textit{``Demonstration Or Protest''} in the SIN'15 dataset. The \textit{``Telephones''} concept was described as {\textit{``speaking on a telephone''}} and {\textit{``talking on a telephone''}}, and its XinfAP went from 0.0 to 0.3151 and from 0.0 to 0.308 in the SIN'13 and SIN'15 datasets, respectively. Similarly, {\textit{``Bicycling''}} which was described as {\textit{``a man riding a bike''}}, {\textit{``people riding bicycles''}} and {\textit{``a woman on a bike''}} achieved 0.373 XinfAP, compared to 0.0569 when only the word ``bicycling'' was used. 

To highlight the performance of our approach, we compare our results with different methods that were designed to solve the SIN task, using training video samples. For the SIN'13 dataset, we compared with the work presented in \cite{MarkatopoulouMM6} and the CERTH participation in the TRECVID SIN task in 2013 \cite{certh2013}. For the SIN'15 dataset, we offer a comparison with the CERTH participation in the TRECVID SIN task in 2015 \cite{certh2015}. Table \ref{mrsc_tbl:2} shows that our approach is very competitive, even though the baselines are explicitly designed for the SIN task. Even though our MRSCs detection approach does not outperform the baseline methods on the SIN task, these results are strong evidence that our approach is suitable for the MRSCs retrieval problem where, as opposed to the TRECVID SIN task used for this evaluation, no training data (annotated visual exemplars) are available for the MRSCs.

\begin{figure}
  \centering
  \includegraphics[width=\linewidth]{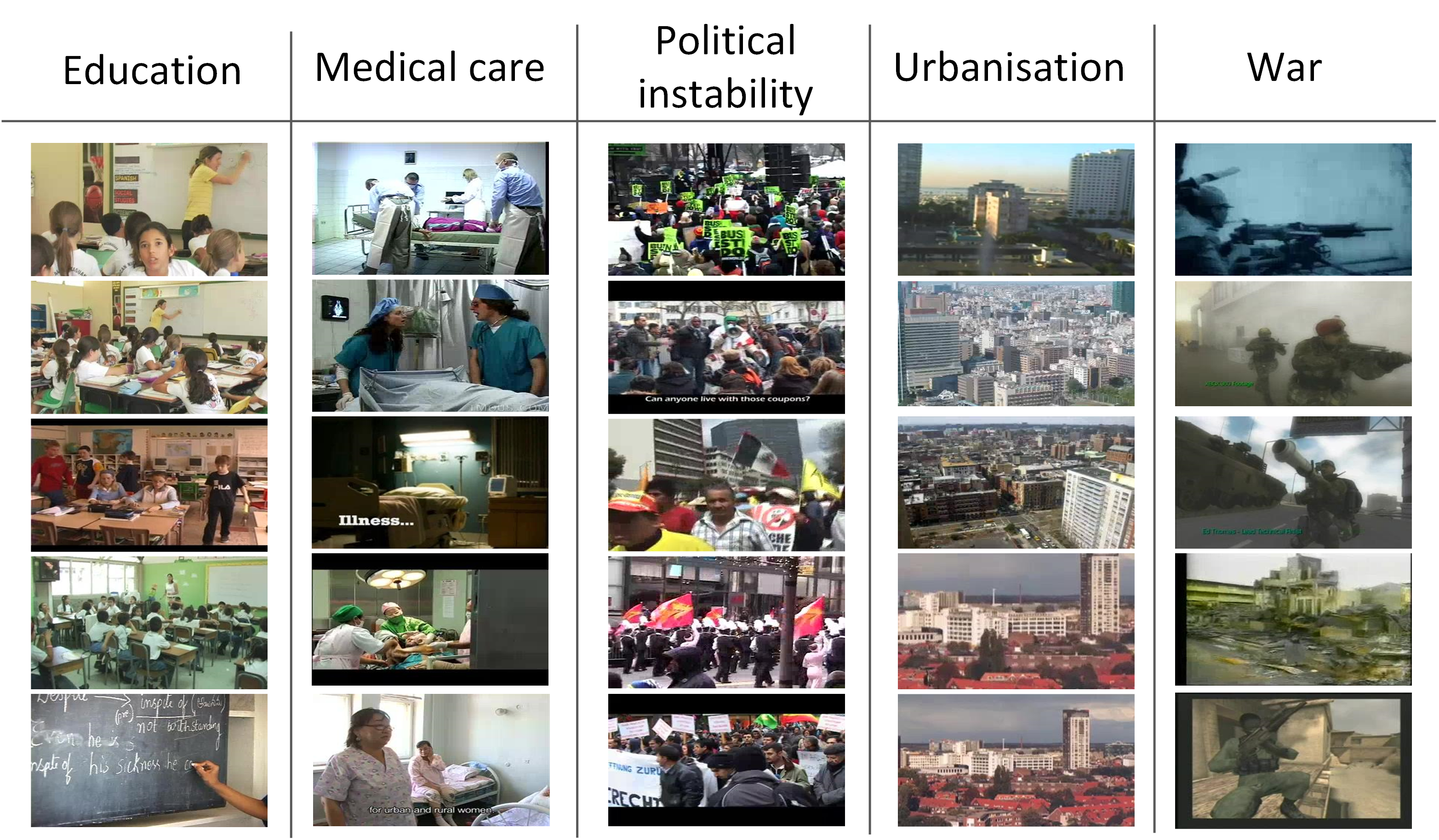}
  \caption{Example results, for five MRSCs (shown on the top of the figure). The top-5 retrieved shots for each MRSC are shown.}
 \label{mrsc_fig:3}
\end{figure}

As mentioned before, there is no MRSC-specific dataset to evaluate our approach's performance empirically. For this reason, we only presented results on the TRECVID SIN datasets. However, to further illustrate our approach's performance, we give some visual examples of the retrieved video shots when we use the MRSCs as input to our method. In Figure \ref{mrsc_fig:3}, the top-5 retrieved shots are presented for a subset of the available MRSCs. These selected MRSCs include representatives of multiple categories and both levels of the hierarchical classification (see highlighted concepts in Table \ref{tab:mrsc}). Although preliminary, results further point to the potential of our approach as a valid and effective solution in finding shots related to MRSCs.

\section{Conclusions and Future Work}
In this paper, we presented an approach that combines two needs for a better understanding of migration decisions and the migration situation: a) a multifaceted view on the migration process, and b) practical automated support for collecting and analyzing relevant video content for this multifaceted perspective. The theoretically founded MRSCs foster a broad view of the migration topic and a common language for analysis. The presented AVS approach can retrieve videos related to abstract MRSC concepts without requiring the time-consuming task of manual video annotation (for training), thus bridging the gap between concepts and video content. The experimental evaluation showed the effectiveness of the proposed approach. Such an approach can be used by border agencies to enrich their analysis of migration contexts and situations with appropriate video coverage. Future research could focus on three aspects: i) fully automated pipeline through automatic MRSCs creation via efficient web harvesting, ii) further performance improvement by AVS method enhancement with better encoding and improved visual and text representations, and iii) further experimentation with domain-specific datasets.

\section*{Acknowledgments} 
This work was supported by the European Union's Horizon 2020 research and innovation programme under grant agreement No 832921 (MIRROR).

%
%
%
\bibliographystyle{splncs04}
 \bibliography{references}
\end{document}